\documentclass[journal,comsoc]{IEEEtran}
\usepackage[left= 0.625in, right= 0.625in,top=0.8in, bottom=0.95in]{geometry}

\usepackage{tablefootnote}
\usepackage{float}
\usepackage{cite}
\usepackage{graphicx}
\usepackage{newtxmath}
\usepackage{makecell}
\usepackage{amsfonts}
\usepackage{amsmath}
\usepackage{algorithm,algcompatible}
\usepackage{fancyhdr}
\usepackage{xcolor}
\usepackage{multirow}
\usepackage{flafter}
\usepackage{subcaption}
\usepackage{url}
\usepackage{enumerate}
\usepackage{epstopdf}
\usepackage{authblk}
\usepackage{enumitem}
\usepackage{booktabs}
\usepackage{array}

\begin{document}
\title{Learning Efficient Communication Protocols for Multi-Agent Reinforcement Learning}

\author{Xinren Zhang, Jiadong Yu, \textit{IEEE Member}, Zixin Zhong
\vspace{-.7cm}
\thanks{X. Zhang, J. Yu, and Z. Zhong are with the Hong Kong University of Science and Technology (Guangzhou).}
}

\maketitle
\begin{abstract}
Multi-Agent Systems (MAS) have emerged as a powerful paradigm for modeling complex interactions among autonomous entities in distributed environments. In Multi-Agent Reinforcement Learning (MARL), communication enables coordination but can lead to inefficient information exchange, since agents may generate redundant or non-essential messages. While prior work has focused on boosting task performance with information exchange, the existing research lacks a thorough investigation of both the appropriate definition and the optimization of communication protocols (communication topology and message). To fill this gap, we introduce a generalized framework for learning multi-round communication protocols that are both effective and efficient. Within this framework, we propose three novel Communication Efficiency Metrics (CEMs) to guide and evaluate the learning process: the Information Entropy Efficiency Index (IEI) and Specialization Efficiency Index (SEI) for efficiency-augmented optimization, and the Topology Efficiency Index (TEI) for explicit evaluation. We integrate IEI and SEI as the adjusted loss functions to promote informative messaging and role specialization, while using TEI to quantify the trade-off between communication volume and task performance. Through comprehensive experiments, we demonstrate that our learned communication protocol can significantly enhance communication efficiency and achieves better cooperation performance with improved success rates.

\end{abstract}

\section{Introduction}
\label{sec:intro}

Multi-Agent Systems (MAS) provide a framework for designing and analyzing distributed autonomous entities that interact within shared environments to achieve individual or collective goals~\cite{wooldridge2009introduction}. Multi-Agent Reinforcement Learning (MARL) extends traditional single-agent Reinforcement Learning (RL) to such multi-agent settings, enabling agents to develop adaptive cooperative or competitive behaviors that effectively model complex systems where single-agent approaches prove insufficient~\cite{sunehag2017value, rashid2020monotonic, foerster2018counterfactual, son2019qtran}. However, this extension introduces significant challenges, including nonstationarity from concurrent learning dynamics, scalability concerns with increasing agent numbers, and the intricacy of achieving effective coordination~\cite{ning2024survey}.

These challenges in MARL are fundamentally rooted in partial observability~\cite{sondik1971optimal}, where agents cannot access the complete system state and thus struggle to determine which information is task-relevant or beneficial to other agents. 
Meanwhile, as the number of agents in future wireless networks grows, the complexity of coordination intensifies significantly, especially when handling high-dimensional data in uncertain environments.
In this context, introducing communication protocols in MARL offers a promising solution by enabling agents to share information to collectively reduce environmental uncertainty~\cite{sukhbaatar2016learning, jiang2018learning, liu2020multi, singh2018learning, das2019tarmac}.

The communication protocol in MAS defines the rules and mechanisms governing agent interactions, encompassing two fundamental dimensions: communication topology (the connectivity structure and temporal coordination among agents) and message content (the information encoded for transmission). Communication protocols can be broadly categorized into three paradigms: engineered, hybrid, and learned approaches.

\textbf{Engineered protocols} rely on technical specifications and predefined structures grounded in rigorous theoretical foundations~\cite{chopra2020evaluation}. Early work established technical methods through state transition models~\cite{bochmann2003formal} and recoverability analysis using Petri Nets~\cite{merlin2003recoverability}, which are later extended to systematic analysis of network control protocols via switched-system approaches~\cite{zou2021communication}. These foundations enable diverse applications including cryptographic authentication for UAV networks~\cite{miao2024uav}, secure routing with intrusion detection~\cite{shifani2024manet}, and event-triggered formation control~\cite{zhou2024interleaved}. While providing reliability and theoretical guarantees, engineered protocols suffer from limited adaptability to dynamic environments.

\textbf{Hybrid approaches} integrate structured frameworks with adaptive elements to address this limitation~\cite{habiba2025revisiting}. Recent advances incorporate higher-level intelligence through communication-centric architectures for LLM-based MAS that balance predefined structures with adaptive capabilities~\cite{yan2025beyond}, and AI/ML-driven semantic protocols enable distributed sensing and intelligent resource allocation in next-generation networks~\cite{strinati2025toward}. Despite these advances, hybrid methods remain fundamentally constrained by their dependence on predetermined rules, unable to fully escape the limitations of their structured foundations.

\textbf{Learned protocols}, as employed in MARL, enable agents to autonomously develop communication strategies through continuous interaction, fundamentally overcoming the rigidity of traditional approaches. Unlike predefined communication protocols, learned approaches enable agents to automatically discover and transmit task-specific representations, selectively encoding only the information from local observations that is relevant to achieving their objectives.
The evolution of learned communication protocols has progressed through several architectural innovations. Early approaches introduced differentiable broadcasting mechanisms~\cite{foerster2016learning,sukhbaatar2016learning}, enabling all agents to share information through gradient-based optimization. Subsequent developments incorporated selective communication via context-dependent gating~\cite{singh2018learning}, allowing agents to dynamically determine when to communicate based on environmental conditions. More recent architectures leverage attention-based message prioritization~\cite{niu2021multi} to weight information relevance, and support heterogeneous agent structures~\cite{seraj2022learning} that enable specialized roles and adaptive coordination. 
This learning-based paradigm adapts dynamically to environmental changes and evolving task requirements, making it particularly suitable for complex, uncertain scenarios where predefined communication rules prove insufficient. 


However, application of MARL algorithms with communication protocols faces critical resource constraints, as real-world limitations in bandwidth, energy, and computational capacity~\cite{10342771} directly impact system feasibility and scalability. Traditional MARL evaluation focuses primarily on task success rates and convergence speed, often overlooking the communication resources required to achieve these outcomes. This gap is problematic as excessive communication overhead can negate the benefits of learned protocols, making communication efficiency a fundamental dimension for transitioning MARL from theoretical research to practical deployment under real-world resource limitations.

Communication efficiency assessment hinges on three key aspects. First, information entropy per successful task completion reveals how effectively agents minimize redundancy while maintaining superior performance~\cite{renyi1961measures}. Achieving objectives with minimal information transfer maximizes the utility of each transmitted bit. We propose the \textbf{Information Entropy Efficiency Index (IEI)} to quantify how effectively agents encode task-relevant information into compact representations.

Second, information diversity during task completion significantly affects system performance~\cite{chen2003performance}. In complex cooperative tasks, heterogeneous roles and specialized sharing lead to superior outcomes, while highly similar information creates redundancy that fails to leverage distributed multi-agent perception. Diverse communication enables agents to partition the information space, with each agent contributing complementary insights for effective division of the task. We propose the \textbf{Specialization Efficiency Index (SEI)} to assess agent differentiation through message diversity.

Third, communication cost-effectiveness reveals how efficiently agents extract value from limited resources~\cite{chafii2023emergent}. This return-on-investment directly impacts scalability and deployment feasibility, as inefficient protocols cause exponential bandwidth growth in large-scale applications. We propose the \textbf{Topology Efficiency Index (TEI)} to quantify task success relative to communication costs.

Together, these three communication
efficiency metrics (CEMs)-IEI, SEI, and TEI-form a comprehensive framework for evaluating communication efficiency in MAS. By incorporating IEI and SEI directly into the policy update process through refined loss functions, agents are guided toward developing more efficient communication protocols, while using TEI as an evaluation metric to assess overall communication cost-effectiveness. This approach addresses a critical gap in current MARL research, where communication efficiency is typically an afterthought rather than an explicit optimization objective.

The contribution of this paper can be summarized as follows:
\begin{itemize}
\item \textbf {Develop} a generalized framework for MARL that accommodates multi-round communication scenarios, providing a unified approach for analyzing diverse MARL architectures.
\item \textbf {Design} three comprehensive metrics (IEI, SEI, and TEI) that quantify different dimensions of communication efficiency in MARL systems, enabling systematic comparison across algorithms.
\item \textbf{Propose} loss function refinements that explicitly optimize $\Phi_{\text{IEI}}$ and $\Phi_{\text{SEI}}$ to enable learning of efficiency-augmented communication protocols, with performance validated through comprehensive $\Phi_{\text{TEI}}$ assessment.
\item \textbf{Conduct} comprehensive experiments demonstrating that our learned communication protocols substantially improve communication efficiency while achieving enhanced cooperative performance.
\end{itemize}
The paper is organized as follows: Section~\ref{Sec2:Related Works} reviews related literature on the learned communication protocol in MARL. Section~\ref{sec:MODEL} introduces our generalized MARL model that accommodates multi-round communication scenarios. Section~\ref{sec: Proposed Metrices} proposes three novel metrics for evaluation of communication efficiency and presents comparative experimental results across different baseline algorithms. Section~\ref{sec: Adjust the loss function} incorporates these metrics into loss functions during policy update to enhance both communication efficiency and task performance. Section~\ref{Sec:EXP} presents the experiment results of the proposed efficiency-augmented method. Finally, Section~\ref{sec: Conclusion} summarizes our contributions and findings.

\section{Related Works} \label{Sec2:Related Works}
The existing literature offers numerous MARL algorithms with learned communication protocols that enable agents to autonomously develop coordination strategies through experience. These learned protocols comprise communication topology and message content. We systematically review these approaches by first examining how different protocol aspects are learned, and next discussing the technical mechanisms that enable this learning, with particular focus on attention mechanisms which is an increasingly prevalent tool for enhancing protocol design.

\subsection{Learned Communication Protocols}
Learned communication protocols in MARL can be categorized based on which aspects of coordination they address: topology control, message content design, or both of them.
\subsubsection{Topology Control}
Communication topology determines the structure and timing of information flow, addressing when agents should communicate and with whom. Foerster et al.~\cite{foerster2016learning} pioneered this direction with RIAL and DIAL, two foundational frameworks that established learnable topology as a component of coordination strategy. RIAL combined deep recurrent Q-networks with independent Q-learning, conditioning each agent's decisions on hidden states, observations, and received messages through cross-timestep broadcasting. DIAL advanced this by enabling gradient flow among agents during centralized learning, using real-valued communication vectors during training that are later discretized during execution.

Recognizing that maintaining permanent communication links may be inefficient, Singh et al.~\cite{singh2018learning} developed IC3Net, which incorporates a gating mechanism that enables agents to learn when communication is necessary across cooperative, competitive, and mixed scenarios. This framework employs independent LSTM controllers with individualized reward signals, utilizing a gating function to modulate the weighted aggregation of other agents' hidden states. This approach represents a significant shift from uniform broadcasting toward selective, context-dependent communication activation. Building upon this concept, SchedNet~\cite{kim2019learning} further advances selective communication by proposing a weight-based scheduler that dynamically determines which agents should broadcast their messages at each time step.

Niu et al.~\cite{niu2021multi} further advanced topology control through MAGIC, which models communication as a learnable directed graph. The framework integrates a scheduler to determine when and to whom to communicate, producing adjacency matrices for targeted message passing through multiple rounds, enabling selective message conveyance and aggregation based on both local and global context.
\subsubsection{Message Content Design}
Message content design determines what information is transmitted and how it is encoded, fundamentally affecting the informativeness and efficiency of inter-agent coordination. The predominant paradigm focuses on hidden state sharing, where communication is realized by broadcasting internal representations.

Sukhbaatar et al.~\cite{sukhbaatar2016learning} introduced CommNet, exemplifying this approach through a neural network framework enabling fully cooperative agents to learn communication alongside their policy. Each agent broadcasts its hidden state as a continuous vector message, with these messages being averaged to form communication input. This end-to-end differentiable system can be trained via standard backpropagation and be combined with RL algorithms for partially observable environments. While simple and effective in small cooperative tasks, hidden state sharing tends to produce redundant signals as agent populations grow, motivating the development of more selective mechanisms that apply additional transformations to hidden states. In HetNet~\cite{seraj2022learning}, each agent generates its hidden state by processing class-specific observations through a tailored module comprising a CNN/fully-connected layer and an LSTM cell. This hidden state is then transformed via a class-specific weight matrix and encoder to form the communication message. In order to enhance bandwidth efficiency, the message can be further binarized using the Gumbel-Softmax function. Finally, receiving agents decode incoming messages and aggregate them using class-adapted attention coefficients, which are subsequently fused with their own hidden states.
\subsubsection{Joint Topology-Content Learning}
Recent approaches address both topology and content simultaneously, recognizing their interdependence in effective communication protocols. Das et al.~\cite{das2019tarmac} developed TarMAC, implementing targeted communication where agents learn both what messages to send (content) and whom to address (topology). Communication messages consist of a signature encoding properties of intended recipients and a value containing the actual message content. The model implements multi-round communication where agents use query vectors to compute attention weights, creating targeted information flow that adapts dynamically to task context.

Liu et al.~\cite{liu2020multi} proposed G2ANet, advancing joint learning through a two-stage design combining hard and soft attention. The framework employs hard attention to identify relevant agent interactions (topology selection) and soft attention to learn relationship importance (message weighting). Using a graph neural network structure, the model determines which interactions to maintain and their relative importance, enabling efficient communication across diverse scenarios.
\subsection{Attention Mechanisms in MARL Communication}
Attention mechanisms have emerged as a key technical approach for improving communication protocols, enabling selective processing of information for both topology control and message design. These mechanisms overcome the inefficiency of indiscriminate broadcasting by introducing selectivity and targeting into learned protocols.
\subsubsection{Attention for Topology Selection}
Attention mechanisms enable dynamic determination of communication partners based on task context. ATOC~\cite{jiang2018learning} employs an attention mechanism to decide if an agent should communicate in its observable field. MAGIC~\cite{niu2021multi} employs attention-based scheduling to produce adjacency matrices that specify directed communication links, allowing agents to selectively determine whom to communicate with at each timestep. G2ANet~\cite{liu2020multi} implements hard attention for topology selection, using binary gating to identify relevant agent interactions while pruning irrelevant connections. This selective activation reduces communication overhead by establishing sparse connectivity patterns that adapt to task requirements.
\subsubsection{Attention for Message Aggregation}
Beyond topology selection, attention mechanisms enable intelligent processing and integration of received messages. TarMAC~\cite{das2019tarmac} implements signature-based soft attention where agents compute attention weights through dot products between query vectors and message signatures, enabling targeted information extraction from multi-agent communications. G2ANet~\cite{liu2020multi} employs soft attention to weight the importance of different agent contributions, processing communication through weighted sums of hidden states. MAGIC~\cite{niu2021multi} employs attention-based message processor to help agents integrate messages for intelligent decision making. This attention-weighted aggregation allows agents to prioritize relevant information while filtering out less pertinent signals, improving both communication efficiency and coordination quality.

Despite these advances in learned protocols and attention mechanisms, evaluation of these frameworks often lacks practical insights for real-world deployment where resource constraints play critical roles. Meanwhile, some approaches employ one-round communication while others implement multiple rounds, highlighting the need for systematic evaluation that considers both communication efficiency and task performance in resource-constrained environments.

\begin{figure*}[t]
  \centering
  \includegraphics[width=.975\linewidth]{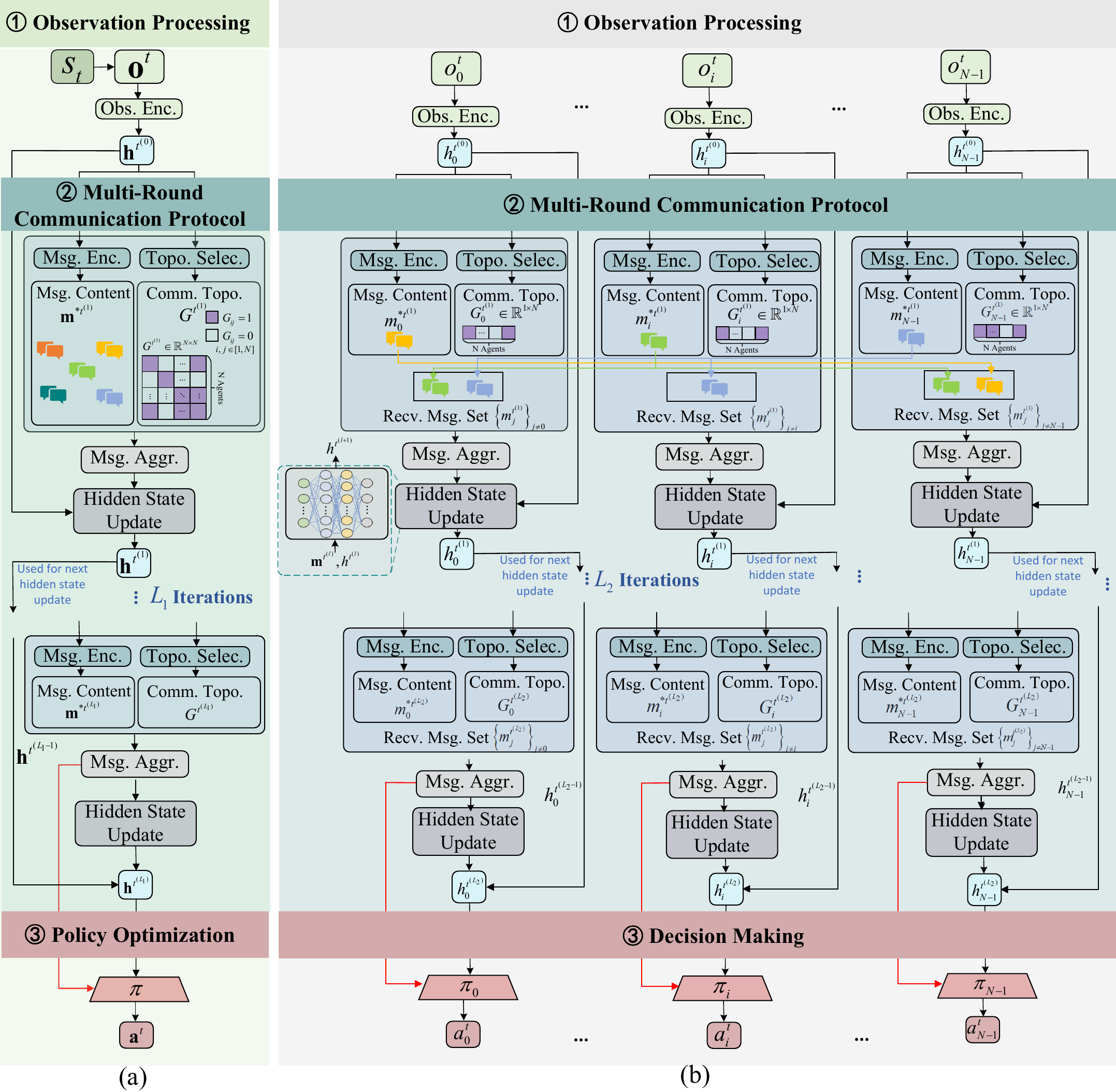}
  \caption{General MARL Framework of Multi-round Communications. (a) Centralized Training, (b) Decentralized Execution.}
  \label{fig:EC_MARL_framework_modeling_with_L-Round_Communications}
\end{figure*}

\section{MARL Framework with Communication Protocol Learning}
\label{sec:MODEL}
To provide a unified formulation for existing MARL communication protocol learning approaches, we propose a generalized framework grounded in a partially observable multi-agent Markov game with an $L$-round communication protocol. We formulate the framework as a tuple $\langle \mathbb{S}, \mathbb{A}, T, \mathbb{O}, \Omega, R, N, \gamma , L\rangle$. Here, $\mathbb{S}$ represents the global state space of the environment, $\mathbb{A}$ encompasses the action space of agents, $T: \mathbb{S} \times \mathbb{A} \rightarrow \mathbb{S}$ denotes the transition function, $\mathbb{O}$ constitutes the observation space of agents, $\Omega: \mathbb{S} \rightarrow \mathbb{O}$ signifies the observation function, $R: \mathbb{S} \times \mathbb{A} \rightarrow \mathbb{R}$ characterizes the reward function, $N$ indicates the number of agents, $\gamma$ is the discount factor, and $L$ counts the rounds of communication protocol.

The proposed generalized model employs the Centralized Training with Decentralized Execution (CTDE) paradigm over multi-round communication. It operates in two distinct stages: centralized training stage and decentralized execution stage. Each stage comprises three fundamental components: (1) observation processing, (2) multi-round communication protocol, and (3) policy optimization or decision making. During the centralized training stage, global information is leveraged to optimize policies and communication mechanisms simultaneously, while in the decentralized execution stage, agents operate independently using only local observations and learned communication protocols. The overall architecture of MARL with $L$-round communications is shown in Fig.~\ref{fig:EC_MARL_framework_modeling_with_L-Round_Communications}.\footnote{Note that $L$ may differ between training and execution: there are $L_1$ communication rounds during centralized training and $L_2$ rounds during decentralized execution, where $L_1 \neq L_2$ in general.}

\subsection{Centralized Training Stage}
During the training phase, global information is leveraged to facilitate more effective learning. As shown in Fig.~\ref{fig:EC_MARL_framework_modeling_with_L-Round_Communications}(a), the centralized training framework consists of the following integrated components:

\subsubsection{Observation Processing}
Each agent $i \in [0,N-1]$ receives its local observation $o_i^t$ from the environment with global state $s_t$. During the centralized training phase, these observations are processed collectively. The collective observations $\mathbf{o}^t = [o_0^t, o_1^t, \ldots, o_{N-1}^t]$ are transformed through a shared observation encoder into initial hidden states:
\begin{equation}
    \mathbf{h}^{t^{(0)}} = f_{\boldsymbol{\theta}_\text{obs}}^o(\mathbf{o}^t),
\end{equation}
where $\mathbf{h}^{t^{(0)}} = [h_0^{t^{(0)}}, h_1^{t^{(0)}}, \ldots, h_{N-1}^{t^{(0)}}]$ represents the collection of all agents' initial hidden states, and $\boldsymbol{\theta}_\text{obs}$ is the parameter set used to encode observations. $\boldsymbol{\theta}_\text{obs}$ is shared across all agents to promote consistency in feature extraction while maintaining the ability to process agent-specific observations.
These initial hidden states serve as the foundation for the subsequent communication rounds. 
\subsubsection{Multi-Round Communication Protocol}
The training process incorporates $L_1$ communication rounds of information exchange among agents before action selection. For each communication round $l \in \{1,2,\cdots,L_1\}$, agents engage in the following steps:

\begin{itemize}
\item Step 1: Message Encoding. Each agent encodes its current hidden state from the previous round into an initial message vector:
\begin{equation}
    \mathbf{m}^{*t^{(l)}} = f_{\boldsymbol{\theta}_\text{msg}}^m(\mathbf{h}^{t^{(l-1)}}),
\end{equation}
where $\mathbf{m}^{*t^{(l)}} = [m_0^{*t^{(l)}}, m_1^{*t^{(l)}}, \ldots, m_{N-1}^{*t^{(l)}}]$ represents the collection of all agents' initial messages. The function $f_{\boldsymbol{\theta}_\text{msg}}^m$ can take various forms such as an identity transform mapping $m_i^{*t^{(l)}} = {h_i}^{{t}^{(l-1)}}$, a linear transformation where $m_i^{*t^{(l)}} = W^l {h_i}^{{t}^{(l-1)}} + b^l$ with learnable weights $W^l$ and bias $b^l$, or an attention-based transform operating on the collective hidden states to capture cross-agent dependencies.

\item Step 2: Communication Topology Selection. The centralized controller determines the communication topology $G^{t^{(l)}} \in \mathbb{R}^{N \times N}$ based on the latest hidden state:
\begin{equation}
    G^{t^{(l)}} = f_{\boldsymbol{\theta}_\text{topo}}^\text{topo}(\mathbf{h}^{t^{(l-1)}}),
\end{equation}
where $G_{i,j}^{t^{(l)}}=1$ indicates that agent $i$ establishes a communication channel with agent $j$. During training, this topology can be predefined based on domain knowledge~\cite{sukhbaatar2016learning}, determined dynamically through an attention mechanism that weighs the importance of different communication links~\cite{singh2018learning,jiang2018learning, niu2021multi}, or learned via a dedicated neural network component that adapts the communication structure according to the current state of the environment and agents~\cite{li2020deep}.

\item Step 3: Message Aggregation. Based on the initial message vectors and the communication topology, the information can be aggregated from connected agents:
\begin{equation}
    \mathbf{m}^{t^{(l)}} = f_{\boldsymbol{\theta}_\text{aggr}}^\text{aggr}(\mathbf{m}^{*t^{(l)}}, G^{t^{(l)}}),
\end{equation}
where $\mathbf{m}^{t^{(l)}} = [m_0^{t^{(l)}}, m_1^{t^{(l)}}, \ldots, m_{N-1}^{t^{(l)}}]$ represents the collection of all agents' aggregated messages. The function $f_{\boldsymbol{\theta}_\text{aggr}}^\text{aggr}$ can implement various aggregation mechanisms such as simple summation where ${m}_i^{t^{(l)}} = \sum_{j \in \mathcal{N}_i^{t^{(l)}}} m_j^{*t^{(l)}}$~\cite{sukhbaatar2016learning, singh2018learning}, graph attention that assigns different weights to messages via ${m}_i^{t^{(l)}} = \sum_{j \in \mathcal{N}_i^{t^{(l)}}} \alpha_{ij}^{t^{(l)}} m_j^{*t^{(l)}}$ with learned attention coefficients $\alpha_{ij}^{t^{(l)}}$~\cite{das2019tarmac,liu2020multi}, and more complex attention-weighted functions that take both message content and graph structure into account to determine the importance of each message within the agent's receptive field~\cite{jiang2018graph, li2020deep,niu2021multi}.

\item Step 4: Hidden State Update. Following message aggregation, each agent updates its hidden state based on its previous hidden state and the aggregated message information:
\begin{equation}
    \mathbf{h}^{t^{(l)}} = f_{\boldsymbol{\theta}_\text{hsu}}^\text{hsu}(\mathbf{h}^{t^{(l-1)}}, \mathbf{m}^{t^{(l)}}),
\end{equation}
where $\mathbf{h}^{t^{(l)}} = [h_0^{t^{(l)}}, h_1^{t^{(l)}}, \ldots, h_{N-1}^{t^{(l)}}]$ represents the updated hidden states for all agents. The function $f_{\boldsymbol{\theta}_\text{hsu}}^\text{hsu}$ can be implemented using recurrent networks~\cite{sukhbaatar2016learning,singh2018learning,niu2021multi}, or other architectures that effectively integrate temporal and communication information~\cite{das2019tarmac,liu2020multi}.
\end{itemize}

This process repeats for $L$ communication rounds. Although the topology generation parameters $\boldsymbol{\theta}_\text{topo}$ remain fixed throughout all rounds, the communication topology $G^{t^{(l)}}$ evolves dynamically across different communication rounds because it takes as input the continuously updated hidden states $\mathbf{h}^{t^{(l-1)}}$ from the previous round. Consequently, while the topology generation mechanism is shared, the actual communication topologies adapt to the evolving information state of the system.

\subsubsection{Policy Optimization}
Upon completion of all $L_1$ communication rounds, the final hidden states can be used to formulate policies and select actions for all agents:
\begin{equation}
    \mathbf{a}^t = f_{\boldsymbol{\theta}_\pi}^{\pi}(\mathbf{h}^{t^{(L_1)}}),
\end{equation}
where $\mathbf{a}^t = [a_0^t, a_1^t, \ldots, a_{N-1}^t]$ represents the joint action of all agents.

During the centralized training phase, we optimize both the individual policies and the communication protocols simultaneously. This is achieved through a centralized critic that has access to global state information:
    $Q(s_t, \mathbf{a}^t) = f_{\boldsymbol{\theta}_Q}^{Q}(s_t, \mathbf{a}^t).$
The training process optimizes the following comprehensive objective function as follows:
\begin{equation}\label{eq: standard loss func}
    \mathcal{L}_t = l_{\mathbf{a}^t} + w_q l_{Q_t},
\end{equation}
where $l_{\mathbf{a}^t}$ is the policy loss and $l_{Q_t}$ is the critic loss.

The network parameters are updated using gradient descent:
\begin{equation}
    \boldsymbol{\theta} \leftarrow \boldsymbol{\theta} - \eta \nabla_{\boldsymbol{\theta}} \mathcal{L}_t,
\end{equation}
where $\boldsymbol{\theta} = \{\boldsymbol{\theta}_\text{obs}, \boldsymbol{\theta}_\text{msg},
\boldsymbol{\theta}_\text{topo},
\boldsymbol{\theta}_\text{aggr}, \boldsymbol{\theta}_\text{hsu}, \boldsymbol{\theta}_\pi, \boldsymbol{\theta}_Q\}$ represents the collective parameters of all network components, and $\eta$ is the learning rate.

This centralized training approach enables agents to learn sophisticated communication protocols and the learned parameters can later be deployed in a fully decentralized manner during execution.
\subsection{Decentralized Execution Stage}
As shown in Fig.~\ref{fig:EC_MARL_framework_modeling_with_L-Round_Communications} (b), during the decentralized execution stage, each agent $i\in [0,N-1]$ operates independently using only local observations and the fixed parameters $\boldsymbol{\theta}$ learned from centralized training. Unlike the training phase where global information is accessible, agents must now rely exclusively on their own observations and received messages.

\subsubsection{Observation Processing}
Each agent $i$ processes its local observation $o_i^t$ independently:
\begin{equation}
    h_i^{t^{(0)}} = f_{\boldsymbol{\theta}_\text{obs}}^o(o_i^t),
\end{equation}
where $\boldsymbol{\theta}_\text{obs}$ represents the fixed parameters of the observation encoder from training.
\subsubsection{Multi-Round Communication Protocol}
The execution phase follows the $L_2$-round communication structure, with each agent operating autonomously without access to global information. For each round $l \in \{1,2,\cdots,L_2\}$:

\begin{itemize}
\item {Message Encoding:} Agent $i$ encodes its hidden state into a message:
\begin{equation}
    m_i^{*t^{(l)}} = f_{\boldsymbol{\theta}_\text{msg}}^m(h_i^{t^{(l-1)}}).
\end{equation}

\item {Topology Selection:} Agent $i$ independently determines its communication neighbors:
\begin{equation}
    G_i^{t^{(l)}} = f_{\boldsymbol{\theta}_\text{topo}}^\text{topo}(h_i^{t^{(l-1)}}),
\end{equation}
where $G_{i,j}^{t^{(l)}}=1$ indicates that agent $i$ establishes a communication channel with agent $j$.

\item {Message Reception and Aggregation:} Agent $i$ receives messages from agents $j$ where $G_{j,i}^{t^{(l)}}=1$, denoted as $\mathcal{N}_i^{t^{(l)}} = \{j \mid G_{j,i}^{t^{(l)}} = 1\}$, and aggregates them:
\begin{equation}
    m_i^{t^{(l)}} = f_{\boldsymbol{\theta}_\text{aggr}}^\text{aggr}(\{m_j^{*t^{(l)}}\}_{j \in \mathcal{N}_i^{t^{(l)}}}, G_i^{t^{(l)}}).
\end{equation}

\item {Hidden State Update:} Agent $i$ updates its hidden state:
\begin{equation}
    h_i^{t^{(l)}} = f_{\boldsymbol{\theta}_\text{hsu}}^\text{hsu}(h_i^{t^{(l-1)}}, m_i^{t^{(l)}}).
\end{equation}
\end{itemize}

All function parameters $\{\boldsymbol{\theta}_\text{msg}, \boldsymbol{\theta}_\text{topo}, \boldsymbol{\theta}_\text{aggr}, \boldsymbol{\theta}_\text{hsu}\}$ remain fixed from training. The communication topology $G_i^{t^{(l)}}$ may vary in rounds as it depends on evolving hidden states.

\subsubsection{Decision Making}
After $L_2$ communication rounds, each agent $i$ independently selects its action:
\begin{subequations}
\begin{align}
    &\pi_i = f_{\boldsymbol{\theta}_\pi}^{\pi}(h_i^{t^{(L_2)}}), \\
    &a_i^t \sim \pi_i,
\end{align}
\end{subequations}
where $\boldsymbol{\theta}_\pi$ represents the fixed policy parameter. This decentralized decision-making integrates both local observations and collective knowledge acquired through multi-round communication, enabling informed action selection without requiring global state information.

In particular, unlike during the training phase, no parameter updates occur during execution. All network components are deployed with their fixed, learned parameters ($\boldsymbol{\theta}_\text{obs}$, $\boldsymbol{\theta}_\text{msg}$,
$\boldsymbol{\theta}_\text{topo}$,
$\boldsymbol{\theta}_\text{aggr}$, $\boldsymbol{\theta}_\text{hsu}$, and $\boldsymbol{\theta}_\pi$, respectively). The centralized critic with parameters $\boldsymbol{\theta}_Q$ is not used during execution, as it is only needed for training, and is not needed during execution. The comprehensive CTDE process for MARL with communication can be found in Algorithm 1.
\begin{algorithm}[!t]
\label{algo:EC-MARL}
\caption{Learning Communication Protocols for Multi-Agent Systems via Centralized Training and Decentralized Execution}
\begin{algorithmic}[1]

\STATE {\bfseries Initialize:} Parameters replay buffer $\mathcal{D}$, parameter set \newline$\boldsymbol{\theta} = \{\boldsymbol{\theta}_\text{obs}, \boldsymbol{\theta}_\text{msg}, \boldsymbol{\theta}_\text{topo}, \boldsymbol{\theta}_\text{aggr}, \boldsymbol{\theta}_\text{hsu}, \boldsymbol{\theta}_\pi, \boldsymbol{\theta}_Q\}$.

\STATE \texttt{/* Phase 1: Centralized Training */}
\WHILE{not converged}
    \STATE Agents collect observation batch $\mathbf{o}^t$ from environment with global state $s_t$.
    \STATE Generate initial hidden states: $\mathbf{h}^{t^{(0)}} = f_{\boldsymbol{\theta}_\text{obs}}^o(\mathbf{o}^t)$.
    
    \FOR{communication round $l = 1$ to $L_1$}
        \STATE Generate the initial message vector $\mathbf{m}^{*t^{(l)}} = f_{\boldsymbol{\theta}_\text{msg}}^m(\mathbf{h}^{t^{(l-1)}})$. 
        \STATE Get the communication topology $G^{t^{(l)}} = f_{\boldsymbol{\theta}_\text{topo}}^\text{topo}(\mathbf{h}^{t^{(l-1)}})$. 
        \STATE Aggregate message $\mathbf{m}^{t^{(l)}} = f_{\boldsymbol{\theta}_\text{aggr}}^\text{aggr}(\mathbf{m}^{*t^{(l)}}, G^{t^{(l)}})$. 
        \STATE Update hidden state $\mathbf{h}^{t^{(l)}} = f_{\boldsymbol{\theta}_\text{hsu}}^\text{hsu}(\mathbf{h}^{t^{(l-1)}}, \mathbf{m}^{t^{(l)}})$. 
    \ENDFOR
    
    \STATE $\mathbf{a}^t = f_{\boldsymbol{\theta}_\pi}^{\pi}(\mathbf{h}^{t^{(L_1)}})$. 
    \STATE Execute $\mathbf{a}^t$, observe rewards $\mathbf{r}^t$ and next state $s_{t+1}$.
    \STATE Store transition in $\mathcal{D}$ and sample mini-batch.
    \STATE $\mathcal{L}_t = l_{\mathbf{a}^t} + w_q l_{Q_t}$ \COMMENT{Compute loss with critic $Q(s_t, \mathbf{a}^t) = f_{\boldsymbol{\theta}_Q}^{Q}(s_t, \mathbf{a}^t)$}.
    \vspace{1mm}
    \STATE Update parameters: $\boldsymbol{\theta} \leftarrow \boldsymbol{\theta} - \eta \nabla_{\boldsymbol{\theta}} \mathcal{L}_t$.
\ENDWHILE

\STATE \texttt{/* Phase 2: Decentralized Execution */}
\FOR{each timestep $t$ during execution}
    \FOR{each agent $i \in \{0,1,...,N-1\}$ \textbf{in parallel}}
        \STATE Receive observation $o_i^t$ and encode: $h_i^{t^{(0)}} = f_{\boldsymbol{\theta}_\text{obs}}^o(o_i^t)$.
    \ENDFOR
    
    \FOR{communication round $l = 1$ to $L_2$}
        \FOR{each agent $i \in \{0,1,...,N-1\}$ \textbf{in parallel}}
            \STATE Generate the initial message $m_i^{*t^{(l)}} = f_{\boldsymbol{\theta}_\text{msg}}^m(h_i^{t^{(l-1)}})$. 
            \STATE Get the communication topology $G_i^{t^{(l)}} = f_{\boldsymbol{\theta}_\text{topo}}^\text{topo}(h_i^{t^{(l-1)}})$. 
        \ENDFOR
        
        \FOR{each agent $i \in \{0,1,...,N-1\}$ \textbf{in parallel}}
            \STATE Receive messages $\{m_j^{*t^{(l)}}\}_{j \in \mathcal{N}_i^{t^{(l)}}}$ where $\mathcal{N}_i^{t^{(l)}} = \{j \mid G_{j,i}^{t^{(l)}} = 1\}$ based on communication topology.
            \STATE Aggregate messages $m_i^{t^{(l)}} = f_{\boldsymbol{\theta}_\text{aggr}}^\text{aggr}(\{m_j^{*t^{(l)}}\}_{j \in \mathcal{N}_i^{t^{(l)}}}, G_i^{t^{(l)}})$. 
            \STATE Update hidden state $h_i^{t^{(l)}} = f_{\boldsymbol{\theta}_\text{hsu}}^\text{hsu}(h_i^{t^{(l-1)}}, m_i^{t^{(l)}})$. 
        \ENDFOR
    \ENDFOR
    
    \FOR{each agent $i \in \{0,1,...,N-1\}$ \textbf{in parallel}}
        \STATE $a_i^t \sim f_{\boldsymbol{\theta}_\pi}^{\pi}(h_i^{t^{(L_2)}})$. 
    \ENDFOR
\ENDFOR

\end{algorithmic}
\end{algorithm}

\section{Proposed Communication Efficiency Metrics} \label{sec: Proposed Metrices}
Communication efficiency represents a critical yet often overlooked dimension in MARL systems. While existing research predominantly evaluates algorithms through task performance, this focus neglects a fundamental trade-off: systems achieving high performance through excessive communication become impractical in large-scale, resource-constrained deployments~\cite{chen2024communication}. The current evaluation framework is without standardized metrics necessary to characterize this performance-efficiency trade-off.

To address this gap, we propose three novel CEMs that explicitly incorporate communication efficiency alongside task performance, providing a comprehensive evaluation framework for multi-agent coordination.
\subsection{Information Entropy Efficiency Index (IEI)}
IEI quantifies the amount of entropy required per unit of success:
\begin{equation}
    \Phi_{{\text{IEI}}_t} = \frac{H_t}{\mathscr{S}_t},
\end{equation}
where $H_t$ is the average message entropy across all agents and communication rounds in epoch $t$:
\begin{equation}
    H_t = \frac{1}{L} \sum_{l=1}^{L} \left( \frac{1}{N} \sum_{i=0}^{N-1} H(M_i^{t^{(l)}}) \right).
\end{equation}
The entropy of each agent's communication message is calculated as:
\begin{equation}
    H(M_i^{t^{(l)}}) = -\sum_{k} p(m_{i,k}^{t^{(l)}}) \log_2 p(m_{i,k}^{t^{(l)}}),
\end{equation}
where $p(m_{i,k}^{t^{(l)}})$ represents the normalized probability distribution of agent $i$'s message components in communication round $l$.

A lower $\Phi_{{\text{IEI}}_t}$ indicates that agents achieve success while exchanging information with lower entropy, suggesting more efficient encoding of task-relevant information. 
\subsection{Specialization Efficiency Index (SEI)}
SEI reflects the degree of functional specialization among agents by measuring the similarity of their communications relative to task success:
\begin{equation}
    \Phi_{{\text{SEI}}_t} = \frac{\xi_t}{\mathscr{S}_t},
\end{equation}
where $\xi_t$ is the average cosine similarity between messages from different agents across all communication rounds in epoch $t$:
\begin{equation}
    \xi_t = \frac{1}{L} \sum_{l=1}^{L} \left( \frac{2}{N(N-1)} \sum_{i=0}^{N-2} \sum_{j=i+1}^{N-1} \cos(\theta_{i,j}^{t^{(l)}}) \right)
\end{equation}
The cosine similarity between messages from agents $i$ and $j$ is calculated as:
\begin{equation}
    \cos(\theta_{i,j}^{t^{(l)}}) = \frac{{m}_i^{t^{(l)}} \cdot {m}_j^{t^{(l)}}}{\|{m}_i^{t^{(l)}}\| \|{m}_j^{t^{(l)}}\|}.
\end{equation}

A lower $\Phi_{{\text{SEI}}_t}$ indicates that agents achieve success while developing more diverse communication patterns, reflecting greater functional specialization. 
\subsection{\textcolor{black}{Topology Efficiency Index (TEI)}}
TEI measures the effective utilization of communication by relating task success to communication frequency:
\begin{equation}
    \Phi_{{\text{TEI}}_t} = \frac{\mathscr{S}_t}{C_t},
\end{equation}
where $\mathscr{S}_t$ represents the success rate at epoch $t$, and $C_t$ denotes the total number of communications. During each epoch $t$, $C_t$ is calculated as:
\begin{equation}
    C_t = \sum_{l=1}^{L} \sum_{i=0}^{N-1} \sum_{\substack{j=0 \\ j \neq i}}^{N-1} G_{i,j}^{t^{(l)}},
\end{equation}
where $G_{i,j}^{t^{(l)}} \in \{0,1\}$ indicates whether agent $i$ communicates with agent $j$ in round $l$ of epoch $t$.

A higher $\Phi_{{\text{TEI}}_t}$ value indicates that agents achieve better task performance with fewer communication acts, indicating greater communication parsimony.
\begin{figure*}[h]
    \centering
    \includegraphics[width=1\textwidth]{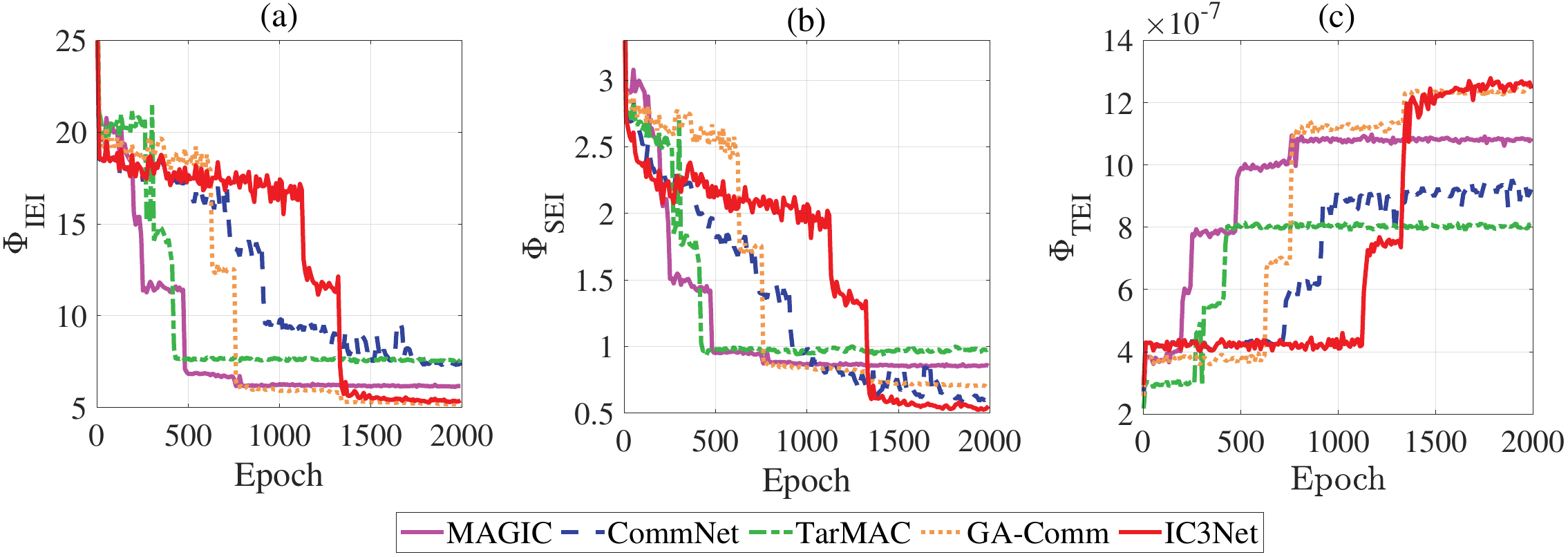}
\caption{Comparison of $\Phi_{{\text{IEI}}}$, $\Phi_{{\text{SEI}}}$, and $\Phi_{{\text{TEI}}}$ for different algorithms in the TJ environment with $L=1$.}  
    \label{fig:CEMs_one_round}
\end{figure*}
\subsection{Comparative Study of CEMs}
\subsubsection{Summary and Comparison of CEMs}
The three proposed CEMs provide complementary perspectives to evaluate communication efficiency in MAS. 
Both the IEI and the SEI capture more implicit aspects of communication efficiency. IEI evaluates the entropy efficiency of transmitted messages, revealing how effectively agents encode task-relevant information into compact representations. SEI, meanwhile, assesses the degree of functional differentiation among agents through message diversity, indicating whether agents have developed specialized communication roles that contribute to coordination.

In contrast, the TEI represents an explicit communication efficiency metric that directly refers to resource consumption, including computational costs, bandwidth utilization, and temporal overhead. By quantifying task success relative to communication frequency, TEI offers a straightforward measure of communication parsimony that aligns with practical deployment constraints in resource-limited environments.

Altogether, these metrics form a comprehensive framework for evaluating communication efficiency in MAS, integrating both implicit message encoding and coordination mechanisms with explicit resource constraints that characterize successful multi-agent learning.
\subsubsection{Experimental Comparison of CEMs}
\begin{figure*}[t]
    \centering
    \includegraphics[width=1\textwidth]{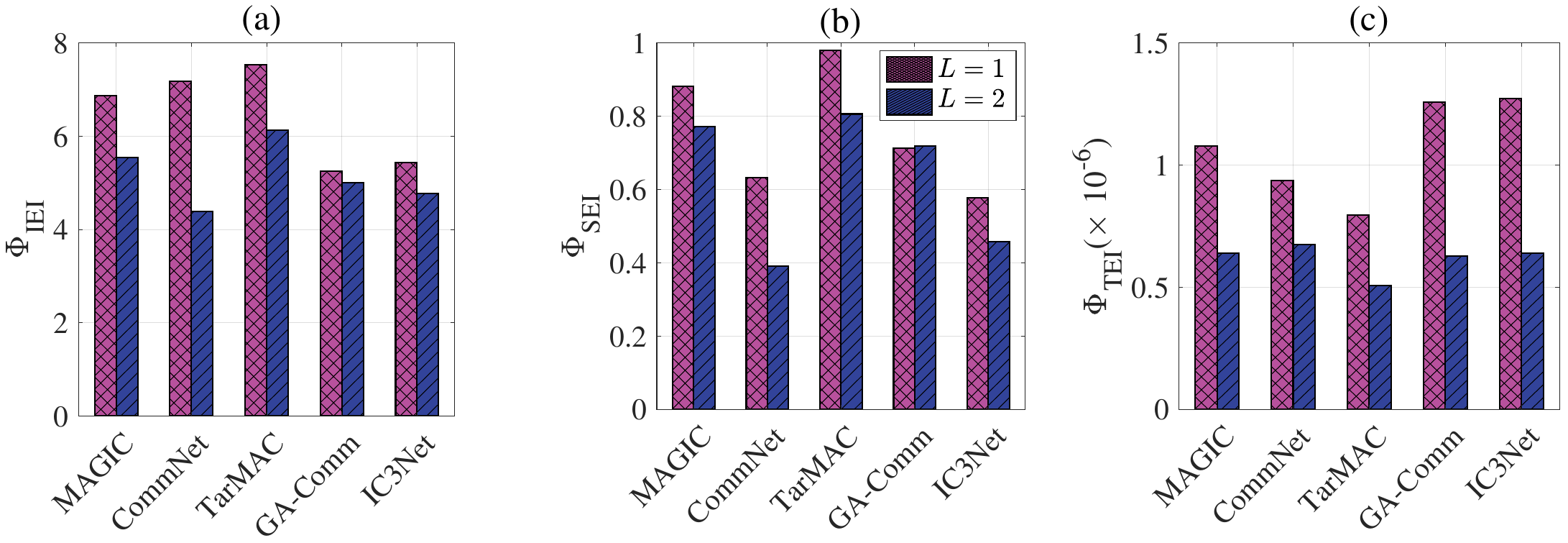}
\caption{Comparison of $\Phi_{{\text{IEI}}}$, $\Phi_{{\text{SEI}}}$, $\Phi_{{\text{TEI}}}$ for different algorithms under one-round and two-round communication scenario in the TJ environment.}  \label{fig:TJ_diff_comm_round_Index123_comparison}
\end{figure*}
Based on the aforementioned metrics, we conduct a comprehensive comparative study of \textcolor{black}{5 MARL algorithms with different learned communication protocols---MAGIC, CommNet, TarMAC, GA-Comm, and IC3Net}---within the Traffic Junction (TJ) environment for $5$ agents~\cite{singh2018learning}. Our evaluation applies  $\Phi_{{\text{IEI}}}$, $\Phi_{{\text{SEI}}}$ and $\Phi_{{\text{TEI}}}$ to systematically assess MAS' communication quality from different dimensions. 
The performance of these 5 algorithms is examined under both one-round ($L=1$) and two-round ($L=2$) communication scenarios, which allows us to study how communication efficiency affects learning dynamics and final performance.
The experiments are conducted using Python 3.10.16 with PyTorch on an Intel (R) Core (TM) i7-14650HX processor and NVIDIA GeForce RTX 4060Ti Laptop GPU.

\begin{table}[htbp]
\centering
\caption{Convergence Epoch and Final Convergence Value Comparison ($L=1$)}
\label{tab:CEMs_convergence}
\renewcommand{\arraystretch}{1.1}
\begin{tabular}{ccccccc}
\toprule
\multirow{2}{*}{\textbf{Algorithm}} & \multicolumn{2}{c}{\textbf{$\Phi_{\text{IEI}}$}} & \multicolumn{2}{c}{\textbf{$\Phi_{\text{SEI}}$}} & \multicolumn{2}{c}{\textbf{$\Phi_{\text{TEI}}$ ($\times 10^{-7}$)}} \\
\cmidrule(lr){2-3} \cmidrule(lr){4-5} \cmidrule(lr){6-7}
& \textbf{Epoch} & \textbf{Value} & \textbf{Epoch} & \textbf{Value} & \textbf{Epoch} & \textbf{Value} \\
\midrule
MAGIC\cite{niu2021multi}    & 750  & 6.5  & 750  & 0.9  & 750  & 10.8 \\
CommNet\cite{sukhbaatar2016learning}  & 1700 & 7.5  & 1700 & 0.6  & 1000 & 9.2  \\
TarMAC\cite{das2019tarmac}   & 400  & 7.5  & 400  & 1.0  & 400  & 8.0  \\
GA-Comm\cite{liu2020multi}   & 1300 & 5.0  & 1300 & 0.75 & 1300 & 12.5 \\
IC3Net\cite{singh2018learning}   & 1300 & 5.0  & 1300 & 0.5  & 1300 & 12.5 \\
\bottomrule
\end{tabular}
\end{table}
Fig.~\ref{fig:CEMs_one_round} presents a comparison of three CEMs for 5 aforementioned MARL algorithms in the TJ environment with one-round communication ($L=1$). Firstly, Fig.~\ref{fig:CEMs_one_round} (a) evaluates $\Phi_{\text{IEI}}$, where lower values indicate more focused communication. All algorithms' $\Phi_{\text{IEI}}$ converge from high initial values to stable low levels, revealing an inverse relationship between convergence speed and communication compactness. TarMAC's $\Phi_{\text{IEI}}$ converges earliest but achieves the highest final value, while IC3Net and GA-Comm's $\Phi_{\text{IEI}}$ converge slower yet attain the lowest value. Secondly, Fig.~\ref{fig:CEMs_one_round} (b) examines $\Phi_{\text{SEI}}$, where lower values reflect more specialized communication. All algorithms reduce initial redundancy following distinct trajectories, exhibiting a similar inverse pattern: TarMAC's $\Phi_{\text{SEI}}$ converges fastest but stabilizes at the highest redundancy, while IC3Net's $\Phi_{\text{SEI}}$ achieves the lowest redundancy despite slower convergence. CommNet and GA-Comm's $\Phi_{\text{SEI}}$ show intermediate performance. Thirdly, Fig.~\ref{fig:CEMs_one_round} (c) compares $\Phi_{\text{TEI}}$, representing integrated efficiency balancing task performance with communication overhead. Unlike IEI and SEI, $\Phi_{\text{TEI}}$ increases during training for all algorithms. TarMAC's $\Phi_{\text{TEI}}$ stabilizes earliest, followed by MAGIC and CommNet's $\Phi_{\text{TEI}}$, with GA-Comm and IC3Net's $\Phi_{\text{TEI}}$ converging last. However, GA-Comm and IC3Net achieve the highest final $\Phi_{\text{TEI}}$, followed by MAGIC, CommNet, and TarMAC.

Across all metrics, a consistent inverse relationship emerges between convergence speed and final performance. IC3Net and GA-Comm converge slowly yet achieve superior values across IEI, SEI, and TEI. Conversely, TarMAC's rapid convergence yields the poorest final values, suggesting premature optimization. MAGIC maintains intermediate performance. These patterns indicate that rapid convergence often signals suboptimal strategy commitment, whereas extended exploration enables more effective task-communication trade-offs. Detailed convergence epochs and final values are shown in Table~\ref{tab:CEMs_convergence}.

Fig.~\ref{fig:TJ_diff_comm_round_Index123_comparison} compares the CEMs for all 5 algorithms under one-round ($L=1$) and two-round ($L=2$) communication scenarios in the TJ environment.
Overall trends reveal a critical trade-off between communication coordination and overhead: while two-round communication achieves superior information focus and specialization (lower $\Phi_{\text{IEI}}$ and $\Phi_{\text{SEI}}$), it incurs substantial communication overhead penalties (lower $\Phi_{\text{TEI}}$) compared to one-round communication. Fig.~\ref{fig:TJ_diff_comm_round_Index123_comparison} (a) demonstrates that two-round communication universally reduces $\Phi_{\text{IEI}}$ relative to one-round scenarios, indicating that additional communication round enables agents to develop more focused information exchange protocols. This improvement is particularly pronounced for CommNet, MAGIC and TarMAC, while GA-Comm and IC3Net show more moderate reductions. Fig.~\ref{fig:TJ_diff_comm_round_Index123_comparison} (b) reveals a similar advantage for two-round communication in specialization. Most algorithms achieve lower $\Phi_{\text{SEI}}$ under $L=2$, reflecting enhanced role differentiation among agents. GA-Comm exhibit comparable performance across both communication rounds. Fig.~\ref{fig:TJ_diff_comm_round_Index123_comparison} (c) presents a contrasting pattern: one-round communication consistently leads to superior $\Phi_{\text{TEI}}$ across all algorithms. GA-Comm and IC3Net experience the largest $\Phi_{\text{TEI}}$ losses under two-round communication (approximately 50\% and 47\% reductions, respectively), while MAGIC and TarMAC show moderate penalties (41\% and 38\%), and CommNet exhibits the smallest degradation (29\%).

The overall findings suggest that while additional communication round enhances encoding and coordination quality, it introduces communication overhead that significantly diminish overall task-communication efficiency in resource-constrained environments.
\section{Protocol Learning with Efficiency Augmentation}\label{sec: Adjust the loss function}

The above findings reveal a critical dilemma: while two-round communication improves message encoding ($\Phi_{\text{IEI}}$) and coordination specialization ($\Phi_{\text{SEI}}$), it severely degrades overall task-communication efficiency ($\Phi_{\text{TEI}}$) due to increased overhead. However, our experiments demonstrate that $\Phi_{\text{IEI}}$ and $\Phi_{\text{SEI}}$ naturally decline during training even with one-round communication, with two-round communication merely accelerating this convergence rather than fundamentally altering it. This observation suggests a key insight that the superior encoding and coordination quality achieved by multi-round communication may be attainable within a one-round framework by explicitly guiding the natural convergence process.
We therefore propose the direct incorporation of $\Phi_{\text{IEI}}$ and $\Phi_{\text{SEI}}$ directly into the training loss function as regularization terms. This approach leverages and accelerates their inherent downward trend, enabling agents to achieve multi-round-level encoding compactness and coordination specialization while maintaining one-round communication. By exploiting this natural learning dynamic rather than adding communication rounds, the MAS obtains superior task-communication efficiency without incurring overhead penalties.

The standard RL training loss function integrated with additional communication efficiency terms can be written as
\begin{equation}\label{eq:loss_adjust}
    \mathcal{L}_t = l_{\mathbf{a}^t} + w_ql_{{Q_t}} + w_{{\text{IEI}}_t}\Phi_{{\text{IEI}}_t} + w_{{\text{SEI}}_t}\Phi_{{\text{SEI}}_t},
\end{equation}
where $\Phi_{{\text{IEI}}_t}$ corresponds to the Information Entropy Efficiency Index, $\Phi_{{\text{SEI}}_t}$ corresponds to the Specialization Efficiency Index, and $w_q$, $w_{{\text{IEI}}_t}$, when $w_{{\text{SEI}}_t}$ are hyperparameters that balance the different objectives. 

\begin{algorithm}
\label{algr2}
\caption{Dynamic Regularization Weight Adjustment for Efficiency-Augmented Training}
\begin{algorithmic}[1]
\STATE {\bfseries Input:} Original loss $\mathcal{L}_t = l_{\mathbf{a}^t} + w_ql_{{Q_t}}$, average message entropy $H_t$ and average message similarity $\xi_{t}$ across all agents and communication rounds in epoch $t$, success rate $\mathscr{S}_t$.
\STATE Set minimum success rate threshold $\mathcal{T}$, scaling factor $\beta$, regularization target ratio $\alpha$, minimum weight $\lambda_\text{min}$, maximum weight $\lambda_\text{max}$, and small constant $\epsilon$.
\IF{$\mathscr{S}_t < \mathcal{T}$}
    \STATE $\mathscr{S}_t \leftarrow \mathcal{T}$.
\ENDIF

\STATE \color{black}{Compute the parameters and metrices in epoch $t$: 
\begin{align*}
    & \Phi_{{\text{IEI}}_t} = H_t \cdot (1.0 - \beta \cdot \mathscr{S}_t),\ \Phi_{{\text{SEI}}_t} = \xi_{t} \cdot (1.0 - \beta \cdot \mathscr{S}_t),
    \\ & \hspace*{-1em}\text{(dynamic weight)} \\& w_{{\text{IEI}}_t} = \alpha \cdot \mathcal{L}_t / (\Phi_{{\text{IEI}}_t} + \epsilon),\ w_{{\text{SEI}}_t} = \alpha \cdot \mathcal{L}_t / (\Phi_{{\text{SEI}}_t} + \epsilon),\\
    &\hspace*{-1em}\text{(Constrain weight range) }\\& w_{{\text{IEI}}_t} = \max(\lambda_\text{min}, \min(\lambda_\text{max}, w_{{\text{IEI}}_t})),\\& 
w_{{\text{SEI}}_t} = \max(\lambda_\text{min}, \min(\lambda_\text{max}, w_{{\text{SEI}}_t})).
\end{align*}}
\STATE Update total loss: $\mathcal{L}_t = \mathcal{L}_t + w_{{\text{IEI}}_t} \cdot \Phi_{{\text{IEI}}_t} + w_{{\text{SEI}}_t} \cdot \Phi_{{\text{SEI}}_t}$.
\STATE {\bfseries Output:} Updated total loss $\mathcal{L}_t$ for efficiency augmentation.
\end{algorithmic}
\vspace{0.2cm}
\noindent{\textbf{Note:} The formulation $\Phi_{{\text{IEI}}_t} = H_t \cdot (1.0 - \beta \cdot \mathscr{S}_t)$ and $\Phi_{{\text{SEI}}_t} = \xi_{t} \cdot (1.0 - \beta \cdot \mathscr{S}_t)$ provides a smoother regularization mechanism compared to the direct division by success rate ($H_t/\mathscr{S}_t$ and $\xi_{t}/\mathscr{S}_t$). This approach reduces sensitivity to success rate fluctuations-applying milder regularization when the success rate approaches 1, while preventing excessive regularization intensity when success rates are low.}
\end{algorithm}
It is noted that while our framework involves three efficiency metrics, we involve merely IEI and SEI in the loss function. This decision stems from both architectural and methodological considerations. Several baseline algorithms in our evaluation (e.g., CommNet, MAGIC) employ fully connected communication topologies where $C_t$ remains constant, making TEI optimization ineffective. Furthermore, while IEI and SEI provide informative gradients about message content and diversity, TEI offers limited training signals for neural network optimization.
Our approach therefore strategically focuses on improving message content efficiency and agent specialization through the learning process using IEI and SEI, while using TEI as an evaluation metric to assess the resource efficiency of communication.

To address the challenge of balancing task performance with communication efficiency in Eq.~(\ref{eq:loss_adjust}), we implement a dynamic regularization weight adjustment mechanism. This approach ensures that communication efficiency is adequately emphasized without compromising the primary objectives of the task.

\begin{table}[t]
\centering
\caption{Simulation parameters}
\label{table2}
\renewcommand{\arraystretch}{1.0}
\begin{tabular}{llll}
\toprule
\textbf{Parameters} & \textbf{Value} & \textbf{Parameters} & \textbf{Value} \\ 
\midrule
$\epsilon$  & $10^{-10}$ & $\mathcal{T}$ & $0.05$ \\
$\beta$ & $0.5$ & $\alpha$ & $0.01$ \\
$\lambda_{\text{min}}$  & $10^{-5}$ & $\lambda_{\text{max}}$  & $5 \times 10^{-3}$ \\
\bottomrule
\end{tabular}
\vspace{-.1cm}
\end{table}
Our dynamic weighting strategy adaptively balances task performance and communication efficiency by modulating regularization intensity based on agent success rates. The mechanism operates with a simple principle: high task performance increases emphasis on efficiency optimization, while performance deterioration automatically reduces regularization pressure to prioritize task completion. This prevents excessive regularization from impeding learning during challenging phases while refining communication protocols during stable periods. The success-rate-dependent scaling maintains proportionality between task objectives and efficiency terms, ensuring optimization stability.
The implementation of our dynamic regularization weight adjustment mechanism is presented in Algorithm 2, which outlines the step-by-step procedure for calculating both IEI and SEI regularization weights based on current performance metrics.

\section{Experiment Results} \label{Sec:EXP}
To validate our proposed efficiency-augmented loss function, we conducted comparative experiments in the TJ environment with $5$ agents. We evaluated the MAS' performance before and after implementing our loss function adjustment. Table \ref{table2} details the key experimental parameters used throughout our evaluation.

\begin{figure}[t]
    \centering
    \includegraphics[width=.445\textwidth]{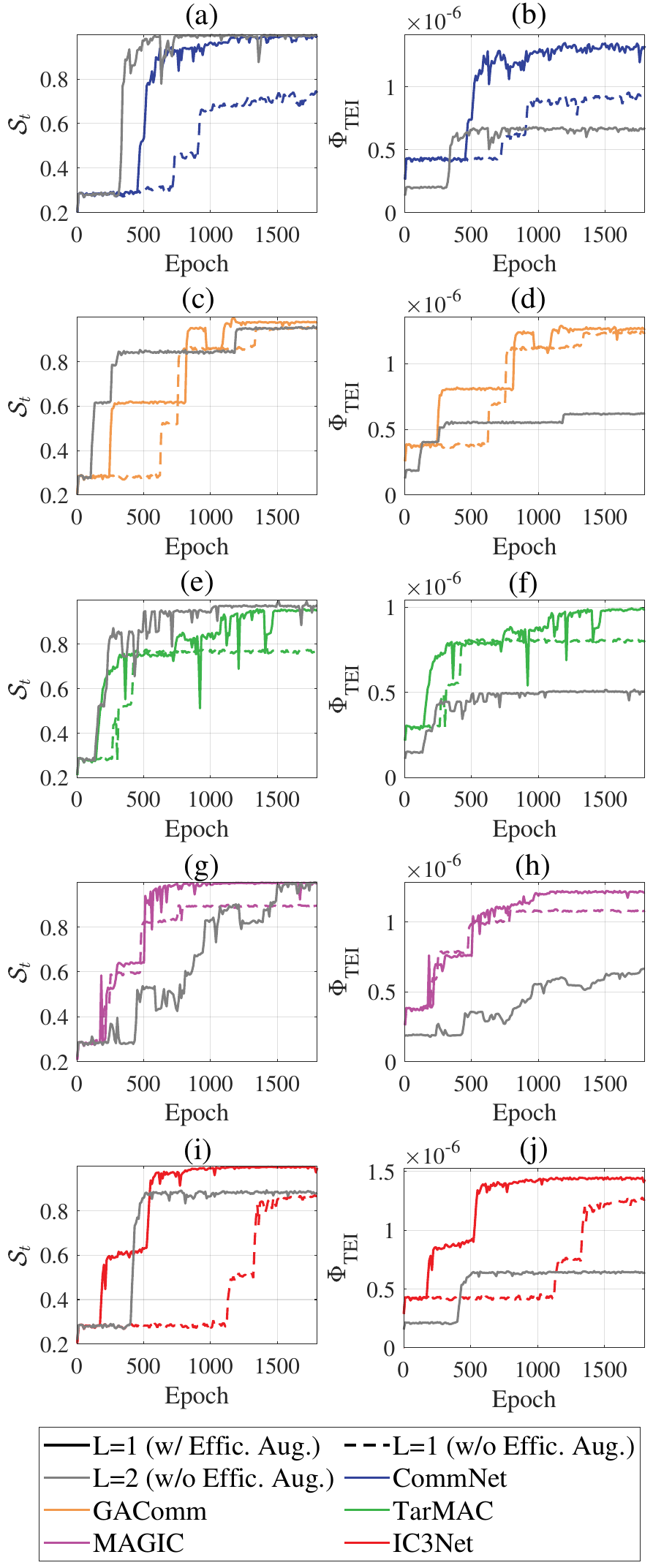}
\caption{Comparison of succcess rate and $\Phi_{{\text{TEI}}}$ for different algorithms in the TJ environment of one-round communication scenario and two-round communication scenario.}  
    \label{fig:TJ_one_round_LA_Success_mean}
\vspace{-.85cm}
\end{figure}

Fig.~\ref{fig:TJ_one_round_LA_Success_mean} presents a comprehensive comparison of 5 communication algorithms under three following experimental conditions:
\begin{itemize}
\item \textbf{$L=1$ (w/o Effic. Aug.):} Baseline configuration with one communication round per decision cycle, using standard training objectives without efficiency augmentation in Eq.~(\ref{eq: standard loss func}).
\item \textbf{$L=1$ (w/ Effic. Aug.):} Configuration with one communication round but enhanced with our proposed efficiency-augmented loss function (Eq.~\ref{eq:loss_adjust}). This explicitly optimizes for both communication efficiency and task performance.
\item \textbf{$L=2$ (w/o Effic. Aug.):} Baseline configuration with two communication rounds per decision cycle, using standard training objectives without efficiency augmentation.
\end{itemize}
The quantitative results based on Fig.~\ref{fig:TJ_one_round_LA_Success_mean} are detailed in Table~\ref{tab:tj_loss_adjust_comparison}.
Across all evaluated algorithms, a consistent pattern emerges: our proposed loss adjustment mechanism simultaneously enhances both task success rate ($\mathscr{S}_t$) and communication efficiency ($\Phi_{\text{TEI}}$). This dual improvement challenges the conventional intuition that performance enhancement necessarily requires increased communication overhead, instead revealing the possibility of synergistic optimization through refined training objectives. We now elaborate more details as below.
\subsection{Efficiency Augmentation in $L=1$}
The 5 algorithms exhibit varying degrees of responsiveness to loss adjustment, which can be categorized into two tiers:
\subsubsection{Strong Responders (IC3Net, CommNet, MAGIC)} IC3Net demonstrates the most transformative response-transitioning from suboptimal plateaus without adjustment to consistently near-perfect performance with adjustment, representing a fundamental capability shift rather than mere acceleration. CommNet shows substantial gains despite its communication-intensive architecture. MAGIC similarly achieves near-perfect $\mathscr{S}_t$ earlier while maintaining higher $\Phi_{{\text{TEI}}}$ values.
\subsubsection{Moderate Responders (TarMAC, GA-Comm)} Both algorithms exhibit meaningful but more modest improvements. TarMAC demonstrates notably greater gains in both $\mathscr{S}_t$ and $\Phi_{{\text{TEI}}}$ compared to GA-Comm, whose improvements are the most limited among all evaluated algorithms.

Although the magnitude of improvement varies, all algorithms demonstrate consistent improvements in both $\mathscr{S}_t$ and $\Phi_{\text{TEI}}$. This universal benefit reveals a key insight: beyond architectural modifications, optimizing the training objective function offers an effective approach to enhance both task performance and communication efficiency. This objective-level strategy is architecture-agnostic, complementing architectural innovations as an alternative pathway to improve task-communication trade-offs in multi-agent learning.
\subsection{Multi-Round Communication}
When comparing $L=1$ (w/ Effic. Aug.) against $L=2$ (w/o Effic. Aug.), algorithm-specific patterns emerge with the following practical implications.
\subsubsection{Superior Performance of One-Round Communication} For GA-Comm, MAGIC, and IC3Net, $L=1$ (w/ Effic. Aug.) consistently outperforms $L=2$ (w/o Effic. Aug.) from the dimensions of both convergence speed and convergence values for $\mathscr{S}_t$ and $\Phi_{\text{TEI}}$ metrics. This indicates that the adjusted communication structure can outperform the increased communication overhead for these architectures.
\subsubsection{Marginal Advantages of Two-Round Communication} CommNet and TarMAC exhibit slightly superior performance with $L=2$ (w/o Effic. Aug.) compared to $L=1$ (w/ Effic. Aug.). However, these modest performance advantages incur substantial efficiency penalties, requiring significantly additional communication overhead to achieve incremental improvements. This trade-off highlights the importance of context-specific optimization strategies.
\begin{table*}[h]
\centering
\caption{Performance Comparison of MARL Algorithms under Different Communication Rounds in TJ Environment}
\label{tab:tj_loss_adjust_comparison}
\renewcommand{\arraystretch}{1.3}
\setlength{\tabcolsep}{1.0pt}
\small
\begin{tabular}{c cccc !{\vrule width 1.0pt} cccc !{\vrule width 1.0pt} cccc}
\toprule
\multirow{3}{*}{\textbf{Algorithm}} 
& \multicolumn{4}{c!{\vrule width 1.0pt}}{\textbf{$L=1$ (w/o Effic. Aug.)}} 
& \multicolumn{4}{c!{\vrule width 1.0pt}}{\textbf{$L=2$ (w/o Effic. Aug.)}} 
& \multicolumn{4}{c}{\textbf{$L=1$ (w/ Effic. Aug.)}} \\
\cmidrule(lr){2-5} \cmidrule(lr){6-9} \cmidrule(lr){10-13}
& \multicolumn{2}{c}{$\mathscr{S}_t$} & \multicolumn{2}{c!{\vrule width 1.0pt}}{$\Phi_{\text{TEI}}$} 
& \multicolumn{2}{c}{$\mathscr{S}_t$} & \multicolumn{2}{c!{\vrule width 1.0pt}}{$\Phi_{\text{TEI}}$} 
& \multicolumn{2}{c}{$\mathscr{S}_t$} & \multicolumn{2}{c}{$\Phi_{\text{TEI}}$} \\
\cmidrule(lr){2-3} \cmidrule(lr){4-5} \cmidrule(lr){6-7} \cmidrule(lr){8-9} \cmidrule(lr){10-11} \cmidrule(lr){12-13}
& Epoch & Value & Epoch & Value 
& Epoch & Value & Epoch & Value 
& Epoch & Value & Epoch & Value \\
\midrule
CommNet\cite{sukhbaatar2016learning}  & 1500 & 0.72 & 1000 & 8.79  & 730 (\textbf{-770}) & 0.99 (\textbf{+.28})  & 730 (\textbf{-270}) & 6.71 (\textbf{-2.08})  & 1200 (\textbf{-300}) & 0.99 (\textbf{+.28}) & 1200 (\textbf{-200}) & 13.47 (\textbf{+4.69}) \\
G2Comm\cite{liu2020multi}   & 1300 & 0.96 & 1300 & 12.35 & 1200 (\textbf{-100}) & 0.95 (\textbf{-.02})  & 1200 (\textbf{-100}) & 6.19 (\textbf{-6.16})  & 1100 (\textbf{-200}) & 0.99 (\textbf{+.03}) & 1100 (\textbf{-200}) & 12.73 (\textbf{+.38}) \\
TarMAC\cite{das2019tarmac}   & 480  & 0.79 & 480  & 8.14  & 1000 (\textbf{+520}) & 0.97 (\textbf{+.18})  & 630 (\textbf{+150}) & 5.04 (\textbf{-3.10})  & 1300 (\textbf{+820}) & 0.95 (\textbf{+.16}) & 1300 (\textbf{+820}) & 9.88 (\textbf{+1.74})  \\
MAGIC\cite{niu2021multi}    & 750  & 0.89 & 750  & 10.88 & 1500 (\textbf{+750}) & 1.00 (\textbf{+.10})  & 1500 (\textbf{+750}) & 6.60 (\textbf{-4.28})  & 700 (\textbf{-50})   & 1.00 (\textbf{+.11}) & 1100 (\textbf{+350}) & 12.22 (\textbf{+1.35}) \\
IC3Net\cite{singh2018learning}   & 1300 & 0.88 & 1300 & 12.39 & 500 (\textbf{-800})  & 0.88 (\textbf{+.01})  & 500 (\textbf{-800}) & 6.36 (\textbf{-6.03})  & 800 (\textbf{-500})  & 1.00 (\textbf{+.12}) & 800 (\textbf{-500}) & 14.48 (\textbf{+2.09})  \\
\bottomrule
\end{tabular}
\vspace{2mm}
\end{table*}
\subsection{Implications for Practical Deployment}
The above findings yield critical insights for deploying MAS under varying communication constraints.
\begin{figure}[t]
    \centering
    \begin{subfigure}{\linewidth}
        \centering
        \includegraphics[width=.99\linewidth]{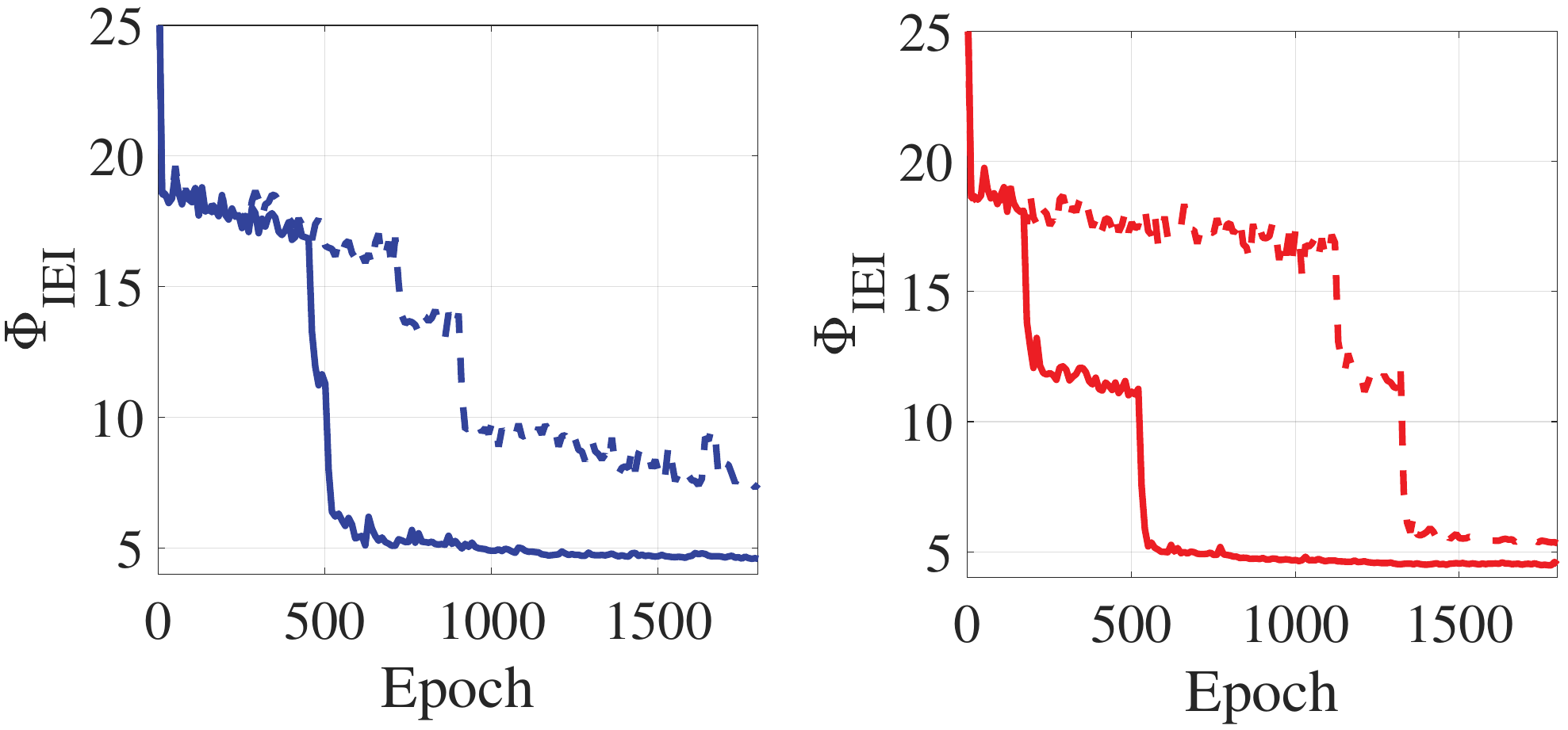}
        \caption{}
        \label{fig:TJ_one_round_LA_IEI_mean}
    \end{subfigure}
    \vspace{0.15cm}
    \begin{subfigure}{\linewidth}
        \centering
        \includegraphics[width=.99\linewidth]{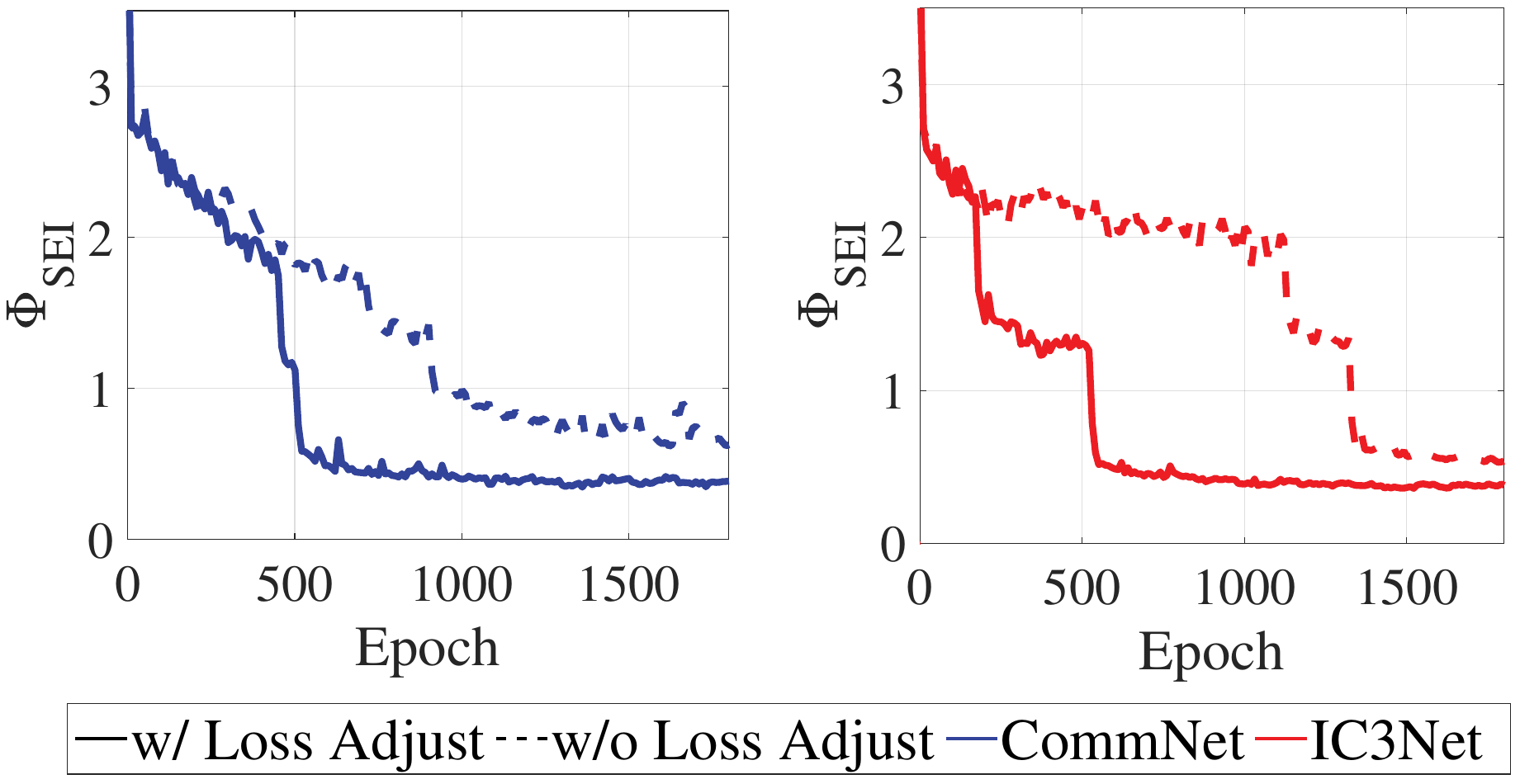}
        \caption{}
        \label{fig:TJ_one_round_LA_SEI_mean}
    \end{subfigure}
    
    \caption{Comparison of (a) $\Phi_{{\text{IEI}}}$ and (b) $\Phi_{{\text{SEI}}}$ for CommNet and IC3Net in the TJ environment of one-round communication scenario.}
    \label{fig:TJ_one_round_comparison}
\end{figure}
\subsubsection{Resource-Constrained Scenarios} When communication constraints are stringent, the better approach is $L=1$ (w/ Effic. Aug.). Because it consistently outperforms it alternative configurations $L=1$ (w/o Effic. Aug.) across all evaluated algorithms. It achieves near-optimal task performance while maintaining significantly enhanced communication efficiency. Therefore,  $L=1$ (w/ Effic. Aug.) enables practical deployment in communication-constrained environments without performance degradation.
\subsubsection{Unconstrained Scenarios} When communication constraints are less stringent, the better approach becomes algorithm-dependent. CommNet and TarMAC marginally benefit from additional communication rounds $L=2$ (w/o Effic. Aug.) in unconstrained settings, while GA-Comm, MAGIC, and IC3Net perform optimally with $L=1$ (w/ Effic. Aug.) regardless of constraint levels.

These results challenge the presumed trade-off between communication efficiency and task performance in MAS. By incorporating efficiency into the optimization objective, our approach enables agents to develop dense communication protocols while maintaining or enhancing task performance.
This reveals that communication inefficiency arises primarily from poorly designed optimization objectives rather than the inherent information needs of the task itself. The consistent patterns observed across architecturally diverse algorithms demonstrate that optimizing objective functions provides a more generalizable solution than modifying network architectures.
\subsection{IEI and SEI Enhancement Comparison}
Fig.~\ref{fig:TJ_one_round_comparison} (a) compares the values of IEI for CommNet and IC3Net with and without our proposed loss adjustment in one-round communication scenarios. Both algorithms show significant efficiency improvements after the implementation of loss adjustment.
The loss-adjusted CommNet model reaches stable IEI values 59\% faster with a 36\% decrease in terminal IEI, while IC3Net shows a 58\% acceleration in convergence with a 19\% reduction in convergence values. These consistent improvements across architecturally distinct algorithms suggest fundamental improvements in information encoding rather than algorithm-specific optimizations.

This reduction in IEI values demonstrates that explicitly incorporating IEI into training objectives leads agents to develop more precise encoding of task-relevant information while minimizing redundant data transmission. This advancement is particularly valuable for MAS with communication constraints. 
GAComm, TarMAC, and MAGIC exhibit similar efficiency patterns, which are not shown due to space limitations.

Fig.~\ref{fig:TJ_one_round_comparison} (b) examines the values of SEI for CommNet and IC3Net with and without our proposed loss adjustment in one-round communication scenarios. Both algorithms demonstrate substantial improvements in communication diversity after implementing loss adjustment.
With loss adjustment, CommNet shows a 48\% acceleration in convergence rate and a 40\% reduction in terminal SEI values, while IC3Net exhibits a 56\% faster convergence with 36\% lower convergence value. These significant reductions in architecturally distinct algorithms suggest that loss adjustment fundamentally transforms communication dynamics rather than optimizing algorithm-specific parameters.

Lower SEI values indicate that agents develop more heterogeneous, specialized communication protocols when efficiency is explicitly incorporated into training objectives. This enhanced diversity enables agents to establish functionally differentiated roles, facilitating more sophisticated coordination while maintaining or improving task performance.
GAComm, TarMAC, and MAGIC still exhibit similar specialized communication patterns, which are not shown due to space limitations.
\section{Conclusions} \label{sec: Conclusion}
This work fills a critical gap in communication research by introducing a framework for evaluating and learning communication protocol efficiency in MAS, particularly focusing on MARL systems. Our proposed CEMs--IEI, SEI, and TEI--provide multidimensional insight into the efficient utilization of information compactness, agent specialization, and communication resources.
Our experiments in various MARL algorithms \textcolor{black}{with different learned communication protocols} imply that traditional approaches often achieve task success at the expense of communication efficiency, while additional communication rounds fail to improve overall efficiency despite potential performance benefits. Most importantly, our refinement of the loss function that incorporates efficiency considerations yields substantial improvements in both communication efficiency and task performance in all evaluated algorithms, demonstrating that these objectives can be optimized simultaneously rather than traded against each other.
As MAS transitions to real-world deployment under practical constraints, our framework provides valuable tools for developing efficient communication protocols. Future work should explore the generalizability of these approaches across diverse environments and communication architectures, including human-agent systems where efficiency constraints are critical.

\bibliographystyle{IEEEtran}
\bibliography{IEEEabrv,mybib}

\end{document}